# Magnetic field-controlled THz modulation in uniaxial anisotropic spin-valves emitters

Arseniy M. Buryakov[1*], Anastasia V. Gorbatova[1], Pavel Y. Avdeev[1], Igor Yu. Pashen'kin[2], Maksim V. Sapozhnikov[1,2,3], Alexey A. Klimov[1], Elena D. Mishina[1], Alexandr S. Sigov[1] and Vladimir L. Preobrazhensky[1]

[1] MIREA – Russian Technological University, Moscow, Russia, 119454
[2] Institute for Physics of Microstructures RAS, Nizhny Novgorod, 603950, Russia
[3] Lobachevsky State University, Nizhny Novgorod, Russia, 603950

[*]buryakov@mirea.ru

Uniaxial spintronic heterostructures constitute compact THz emitters under femtosecond optical excitation, with emission amplitude and polarization governed by the applied magnetic field is presented. We demonstrate here efficient magnetically tunable THz amplitude control in ultrathin, exchange-biased Co/Pt/Co/IrMn spin valve. Terahertz spintronic magnetometry resolves reversible switching between parallel and antiparallel magnetization states and correlates the high- and low-emission regimes with constructive and destructive interference of charge transients generated by the inverse spin Hall effect in the Pt spacer. A residual low-emission signal is traced to spin-to-charge conversion in IrMn. Phase inversion under front- versus back-side excitation confirms the ISHE origin of the emission, while a macrospin Landau-Lifshitz-Gilbert model reproduces the field dependence of its amplitude and separates layer-specific contributions. Together, these results define a wafer-compatible spin valve device architecture that enables efficient, low-field THz amplitude control.

## I. INTRODUCTION

Terahertz (THz) radiation, spanning 0.1–10 THz, is a strategically important spectral range for next-generation wireless communications (6G+) [1,2], biomedical imaging [3,4], spectroscopy [5,6], and emerging quantum technologies [7,8]. Despite its enormous technological promise, the realization of practical THz systems is constrained by fundamental bottlenecks. Active amplitude, frequency, and polarization modulators remain insufficiently developed for on-chip integration, limiting the full potential of THz technology – particularly in secure communications and quantum information processing, where the manipulation of single photons is fundamentally essential [9,10].

Following the discovery of THz generation in ferromagnetic (FM)/non-magnetic metal (NM) bilayers in 2013 [11], interest in these structures as active THz system components has grown significantly. Subsequent research has demonstrated that they serve as efficient THz sources (THz field strengths of ~1.5 MV/cm [12]) with extensive room-temperature modulation capabilities, including amplitude [13,14], phase [15], polarization [16-21] and frequency [22] control directly at the source. Device operation relies on ultrafast spin transport and the inverse spin Hall effect (ISHE) and proceeds through four consecutive stages: femtosecond optical excitation of spin-polarized carriers in the FM layer, their superdiffusion into the NM layer, ISHE-mediated conversion of the transverse spin current into a longitudinal charge current, and emission of a THz pulse.

Spin valves are more complex multilayer structures consisting of two FM layers separated by a NM metal spacer (FM/NM/FM). A key feature of spin-valve architecture is that one FM layer must be magnetically hard (with a pinned magnetization direction, for example, through exchange coupling to an antiferromagnetic layer), while the second must be magnetically soft (free). This configuration provides the foundation for GMR. In THz emitters based on spin valves, this effect can be utilized for amplitude modulation of THz signals. Antiparallel alignment of magnetic moments in the FM layers leads to the summation of charge currents generated via the ISHE in the NM layer, increasing the THz radiation amplitude, whereas parallel alignment results in partial cancellation and reduced THz output. One of the first works on spin valves as THz emitters was presented in Fe(3 nm)/Pt(4 nm)/Fe(3 nm)/IrMn(20 nm) structures [13]. A different concept was proposed in [23], where paired FeCo/$TbCo_2$/FeCo trilayers separated by a copper spacer act as two optically interfered THz sources; magnetic-field control arises from different coercivities, and the final amplitude is determined by interference of THz waves rather than simple current summation.

Despite impressive modulation results from both approaches, certain limitations persist. The relatively large total metallic thickness (≈ 30 nm [13], ≈ 20 nm [23]) in THz spin valves increases optical and THz radiation losses. Contemporary research demonstrates that the optimal layer thickness is approximately 2 nm [24,25]. Furthermore, achieving high efficiency requires careful control of interfaces: FM/NM interface quality plays a critical role in determining spin-to-charge conversion efficiency [26,27]. Magneto-optical effects are likewise interface- and granularity-sensitive in Co-based systems [28], emphasizing the coupling between structure and functionality in FM/NM structures.

Consequently, special attention must be paid to material aspects and interface engineering. The choice of FM and NM layer materials significantly affects THz generation efficiency [24]. Simultaneously, NM spacer thickness in spin valves influences the interlayer exchange interaction between neighboring FM layers and requires careful optimization [14]. Another important factor is substrate selection, which can significantly influence the functional film and interface structure, magnetic anisotropy, and thermal properties of the emitter [25,29,30].

This work investigates ultrathin spin-valve structures for efficient THz generation and amplitude modulation with controlled polarization. We explore Co/Pt/Co/IrMn stacks and address: *i*) optimization of Pt spacer thickness to tune interlayer exchange interaction and spin-to-charge conversion, including substrate effects (sapphire, silicon, amorphous glass); *ii*) quantitative assessment of amplitude modulation depth and field-driven THz polarization changes; and *iii*) theoretical analysis of the physical mechanisms governing THz generation: evaluation of interlayer exchange interactions, layer-resolved (pinned/free FM) current contributions in Pt and IrMn.

## II. MATERIALS AND METHODS

A series of multilayer thin-film stacks with the structure Co (1.8 nm)/Pt (3 nm)/Co (1.8 nm)/IrMn (5 nm) (Fig. 1) was fabricated using magnetron sputtering in an AJA International ATC-2200 system. Deposition was carried out at room temperature with a base pressure of $10^{-7}$ Torr. The FM layers, with a fixed thickness of 1.8 nm, were deposited from the Co target. Platinum was selected as the NM spacer for its large spin Hall angle, which promotes efficient spin-to-charge conversion. Its thickness of 3 nm was optimized to mediate the interlayer interaction and enable a robust spin-valve effect. To investigate substrate-dependent effects, we employed three THz-transparent substrates: glass, sapphire ($\alpha$-$Al_2O_3$), and high-resistivity silicon. A detailed analysis of the effects of Pt thickness and substrates on the magnetic characteristics is provided in the Supplemental Material (Figs. S1 and S2) [31].

To ensure smooth film growth, a 3 nm amorphous Si seed layer was deposited in-situ on all substrates prior to the active stack. Post-deposition, a 3 nm Si capping layer was applied to prevent oxidation. A uniaxial magnetic anisotropy was induced in the FM layers by applying a static in-plane magnetic field of ≈ 180 Oe during the deposition. This also served to set the exchange bias at the FM/IrMn interface. The sputtering parameters were as follows: argon working pressure of $2 \times 10^{-3}$ Torr for the Co, Pt, and Si layers, and of $4 \times 10^{-3}$ Torr for the IrMn layer; deposition rates and RF powers of 0.86 Å/s and 150 W for Co, 2.22 Å/s and 150 W for Pt, 0.33 Å/s and 55 W for IrMn, and 0.21 Å/s for Si.

Magnetic properties of the fabricated spin valves were characterized by the longitudinal magneto-optical Kerr effect (MOKE). THz emission characteristics – amplitude and polarization – were measured using a time-resolved THz spectroscopy setup in transmission geometry. Samples were excited with a femtosecond amplified Ti:Sapphire laser (Avesta Project Ltd.) with a central wavelength of 800 nm, a pulse duration of 45 fs, and a repetition rate of 3 kHz. The pump beam (3 mm spot diameter) was incident normally on the sample surface. Pump fluence varied between 0.014 and 0.471 mJ/cm² using a half-wave plate placed in front of a Glan-Taylor polarizer. An IR cut-off filter after the sample suppressed residual pump light. Emitted THz radiation was detected via electro-optic sampling (EOS) method. Two off-axis parabolic mirrors collected and collimated the THz beam onto a ZnTe EOS detector. The probe beam was focused precisely onto the same point on the ZnTe crystal as the THz beam. Both beams were polarized parallel to the [–110] axis of ZnTe to maximize sensitivity to the $E_X$-component of the THz field. Samples were mounted on a rotating, nonmagnetic stage inside the gap of an electromagnet, which allowed us to investigate magnetic anisotropy and field-induced THz amplitude modulation. A wire-grid polarizer (WGP) was positioned immediately before the ZnTe analyzer to selectively resolve the polarization state of the THz radiation.

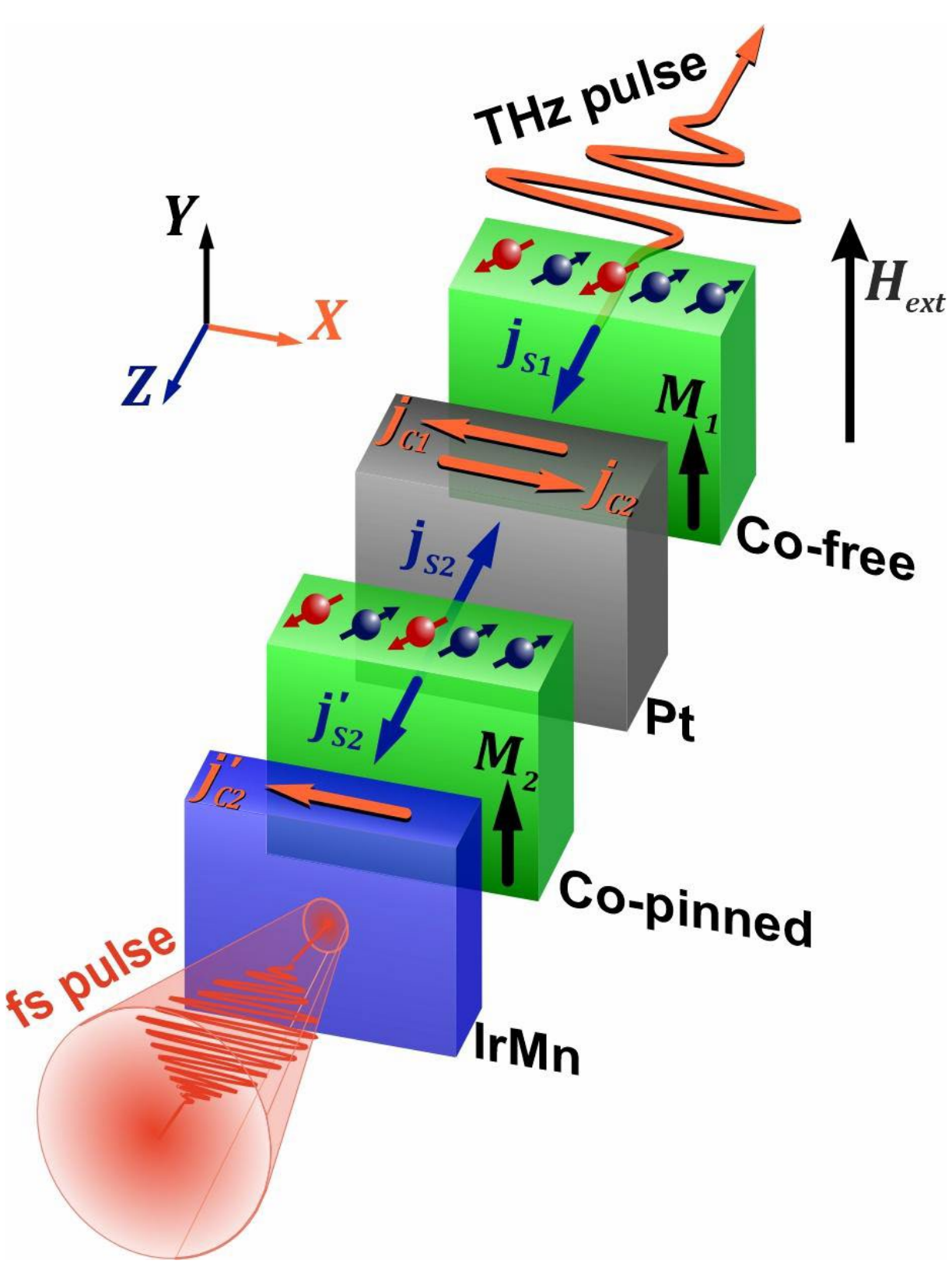

FIG 1. Multilayer structure of the Co (1.8 nm)/Pt (3 nm)/Co (1.8 nm)/IrMn (5 nm) spin-valve with free and pinned FM layers. Excitation by a femtosecond laser pulse launches spin currents $j_{s1}$ and $j_{s2}$ from the free and pinned Co layers, respectively, into the Pt spacer, while an additional spin current $j'_{s2}$ diffuses from the pinned Co into the IrMn layer. The strong spin-orbit coupling in Pt and IrMn converts these spin currents into transverse charge currents via the ISHE: $j_{s1}$ and $j_{s2}$ generate $j_{c1}$ and $j_{c2}$ inside Pt, whereas $j'_{s2}$ produces $j'_{c2}$ within IrMn. The generated THz pulse arises from the superposition of $j_{c1}$, $j_{c2}$, and $j'_{c2}$.

## III. RESULTS

### A. Magneto-optical analysis and THz amplitude modulation

Figs. 2(a) and 2(b) display MOKE hysteresis loops measured for the Co (1.8 nm)/Pt (3 nm)/Co (1.8 nm)/IrMn (5 nm) stack deposited on a high-resistivity silicon substrate. The red and dark-blue loops were measured under in-plane magnetization with the external field $H_{ext}$ applied parallel to the magnetic anisotropy axis (easy axis, EA) and, respectively, along the orthogonal hard axis (HA). The EA loop (Fig. 2(a)) exhibits the characteristic two-step switching behavior of the spin-valve comprising the magnetically soft "free" and exchange-pinned Co layers. The two-step loop occurs because the free layer undergoes magnetization reversal at the low field $H_1 \sim 0.05$ kOe, whereas the pinned layer must overcome both its intrinsic anisotropy and the interfacial exchange coupling, so its reversal occurs only at the highest magnetic field $H_2 \sim 0.55$ kOe. Exchange interaction at the interface with the antiferromagnet causes a shift $H_{EB} \approx 0.565$ kOe in the hysteresis loop of the fixed magnetic layer (exchange bias effect). When magnetized along the EA, this results in two separate loops on the magnetization curve.

Conversely, when magnetized perpendicular to the EA (Fig. 2(b)), the free layer magnetizes in weaker fields, while the pinned layer only saturates in significantly higher fields. As a result, a characteristic kink is observed on the magnetization curve at $H_1 \approx \pm 0.05$ kOe. In the low-field range $0 < H_{ext} \leq H_1$, the normalized projection of magnetization $M/M_s$ rises steeply, reflecting a field-driven 90° rotation of the magnetic moment in the free layer from the EA toward the HA, which completes at the anisotropy field $H_{a1} \approx 0.05$ kOe. Beyond the $H_{ext} > H_{a1}$ the free layer is fully saturated along the HA, and the subsequent, more gradual rise in $M/M_s$ originates from the pinned layer. Stabilized by exchange coupling to IrMn, the magnetic moment of the pinned layer saturates only at $H_{a2} \approx 0.8$ kOe.

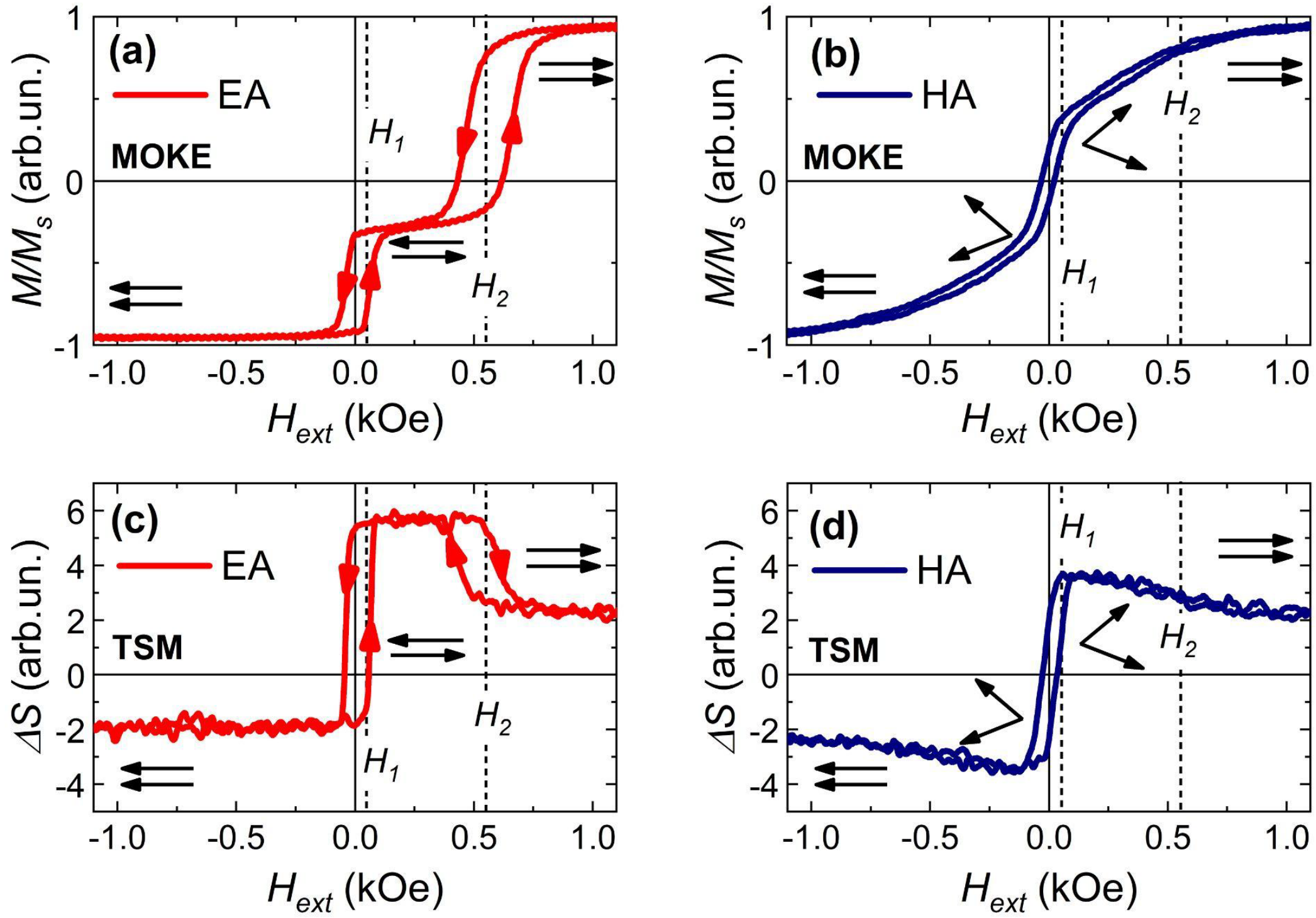

FIG 2. (a),(b) MOKE and (c),(d) THz-TSM characterization of the Co (1.8 nm)/Pt (3 nm)/Co (1.8 nm)/IrMn (5 nm) spin-valve under an in-plane external magnetic field $H_{ext}$. The red and blue curves represent magnetization reversal along the EA and HA, respectively. Paired black arrows at selected regions of the loops schematically depict the relative orientations of the magnetization vectors in the free and exchange-pinned Co layers. Dashed lines outline the loop segments where the free and pinned layers switch upon increasing $H_{ext}$ (at $H_1$ ~ 0.05 kOe and $H_2$ ~ 0.55 kOe, respectively).

Subsequent switching between parallel and antiparallel magnetization states, detected by the MOKE experiment, was independently corroborated by THz spintronic magnetometry (TSM). TSM quantifies the dependence of the peak THz amplitude on the applied magnetic field in the THz-TDS setup. The corresponding data in Figs. 2(c) and 2(d) were obtained for film-side excitation at a fluence of 47.2 μJ/cm$^2$, which is below the threshold where the sample would be heated above the blocking temperature of the antiferromagnet [32]. The TSM hysteresis curves, as in the MOKE experiment, were obtained with the external magnetic field applied either along the in-plane EA and HA (red and dark blue curves in Figs. 2(c) and 2(d), respectively). When magnetized along the EA, the free layer exhibits a narrow, low-coercivity hysteresis loop, whereas the pinned layer displays a broader, exchange-biased loop. The corresponding switching fields are $H_1 \approx 0.05$ kOe for the free layer and $H_2 \approx 0.55$ kOe for the pinned layer, in agreement with values obtained from MOKE measurements. Moreover, in the HA geometry, the analysis of the inflections on the curve yields identical anisotropy fields for the free and pinned layers, 0.05 kOe and 0.8 kOe, respectively.

TSM provides complementary information beyond conventional magnetometry by directly probing the spin-to-charge conversion efficiency within the spin-valve. Unlike MOKE, which monitors static magnetization alignment, TSM reflects the dynamic interference of ultrafast charge currents generated via the ISHE in the NM layers. The relative magnetic configuration of the free and pinned layers determines whether these contributions interfere constructively or destructively, enabling field-controlled amplitude THz modulation. As evident from Fig. 2(c), under antiparallel alignment of the ferromagnetic moments (indicated by the arrows), the THz signal reaches its maximum – approximately 2.5 times higher than in the parallel configuration.

The enhanced THz output arises from the constructive superposition of the charge currents from both FM layers within the NM spacer. As depicted schematically in Fig. 1, femtosecond optical excitation launches spin currents from the free and pinned Co layers into the Pt spacer ($j_{s1}$ and $j_{s2}$), where they are converted into charge transients ($j_{c1}$ and $j_{c2}$) via the ISHE. In the antiparallel configuration, these two charge components add constructively, resulting in the maximum THz emission amplitude. In the parallel state, however, their contributions interfere destructively, which strongly suppresses the THz output. A notable distinction from earlier reports [13] is that the cancellation is incomplete because an additional spin current $j'_{s2}$ diffuses from the pinned Co layer into the adjacent IrMn layer with strong spin-orbit coupling and generates a charge current ($j'_{c2}$) (Fig. 1). Control measurements presented in the Supplemental Material (Fig. S4) [31] confirm that the spin Hall angles of Pt and IrMn have the same sign, which means that $j_{c2}$ and $j'_{c2}$ partially compensate each other (Fig. 1), leading to incomplete suppression of the THz signal.

The experimental data in Fig. 3(a) provide direct evidence supporting the ISHE-driven THz generation mechanism illustrated in Fig. 1. When the excitation beam is incident from opposite sides of the spin-valve grown on the sapphire substrate, the THz signals display a clear phase inversion at a fixed external magnetic field $H_{ext}$ = -2 kOe, while their amplitudes remain comparable. A temporal shift of the waveforms ~ 1.9 ps in Fig. 3(a) originates from the different propagation velocities of the optical pump and the emitted THz wave inside the substrate, determined by its refractive indices at 800 nm and in the THz range. A control experiment performed on an identical spin-valve grown on the glass showed the same phase inversion behavior (Fig. S3 in the Supplemental Material) [31]. By contrast, such two-side excitation tests cannot be carried out for Si substrates at 800 nm due to their strong absorption and low transparency at the pump wavelength.

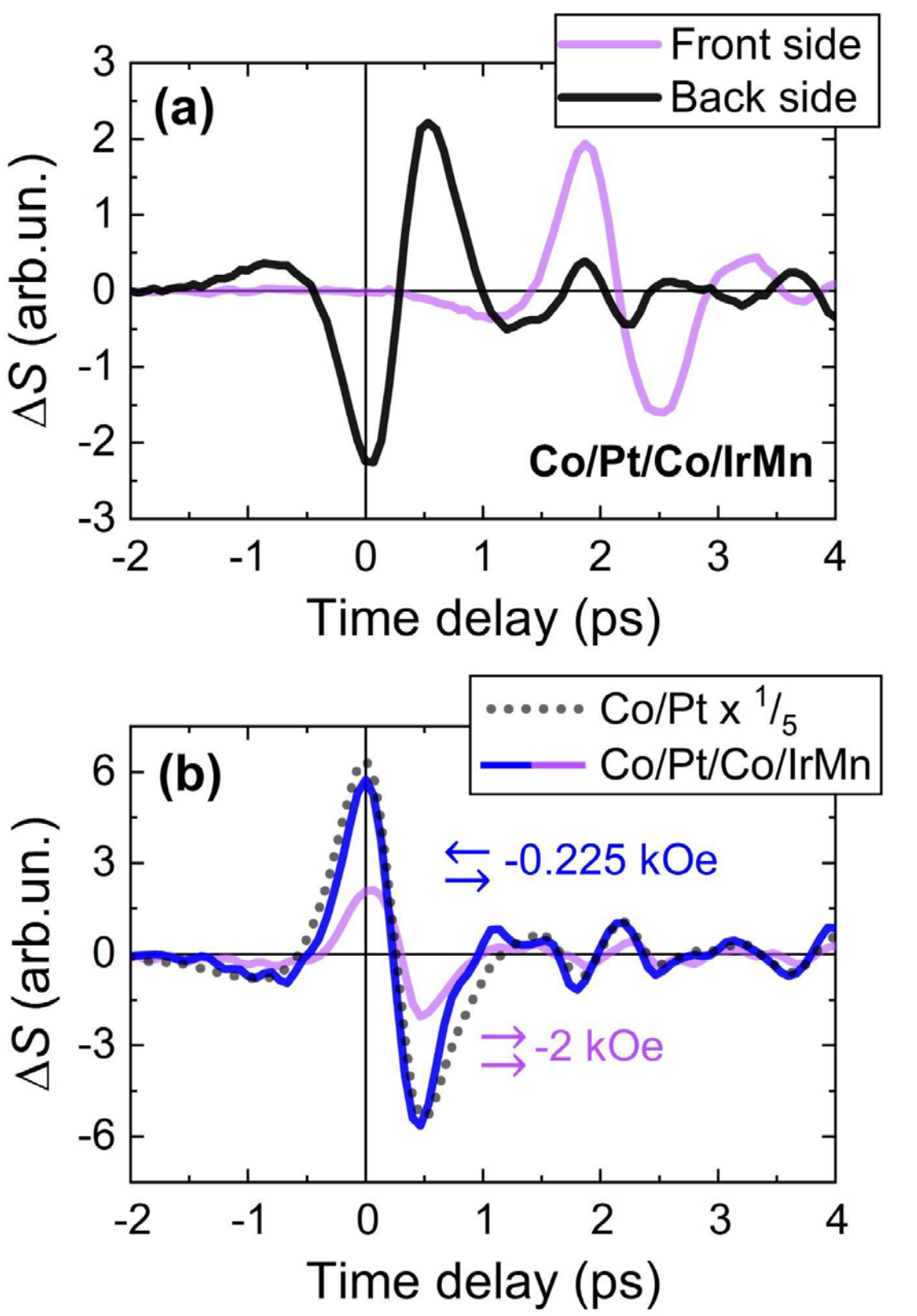

FIG 3. Time-domain THz emission from Co/Pt/Co/IrMn spin valves. (a) THz waveforms from the spin-valve on the sapphire substrate. Reversing the excitation side (front vs back) produces a phase inversion at a fixed external field of $H_{ext}$ = -2 kOe. (b) Relative efficiency of THz emission from the spin-valve on the silicon substrate compared with the Co (2 nm)/Pt (2 nm). Blue and purple curves for the spin-valve correspond to the states with antiparallel and parallel alignments of the magnetic moments of the FM layers, respectively, with the external magnetic field values indicated for each configuration.

Despite the residual THz emission in the antiparallel state, the amplitude-modulation depth in our spin-valve – defined as the relative decrease in THz amplitude when transitioning from the high-amplitude state (taken as 100 %) to the low-amplitude state – reaches ≈ 65.5 %. Although this value is lower than the 93.6% modulation depth reported in Ref. [13], our structure achieves comparable field-controlled THz amplitude tuning with a total metallic thickness reduced to 11.6 nm, mitigating optical and THz losses.

To assess the THz emission efficiency, the spin valve was benchmarked against a Co(2 nm)/Pt(2 nm) bilayer reference emitter measured in the same THz-TDS setup (Fig. 3(b)). In the high-amplitude state (blue curve in Fig. 3(b)) the spin-valve THz field is reduced by only a factor of five relative to the Co/Pt reference, whereas in the low-amplitude state the reduction is about one order of magnitude. Co/Pt bilayers with nominally similar thicknesses, deposited in the same sputtering system under identical conditions, were previously absolutely calibrated and yielded a THz pulse energy fluence of 90 nJ/cm$^2$ at the pump fluence of 1 mJ/cm$^2$, corresponding to the optical-to-THz conversion efficiency of ~$7\times10^{-3}$ [33]. Based on this reference, the optical-to-THz conversion efficiency of the present spin-valve in the high-amplitude state is expected to be of the same order of magnitude. Beyond its enhanced conversion efficiency, our spin-valve exhibits reversible sequential switching among multiple stable magnetic states. Fig. 4(a) displays the TSM hysteresis loop measured at the pump fluence of 0.47 mJ/cm$^2$, with points 1–5 marking representative field values $H_{ext}$ at which subsequent analyses of amplitude modulation and THz polarization were performed. The temporal stability of the peak THz amplitude upon toggling among these field values is corroborated in Fig. 4(b): the emission remains stable and highly reproducible over repeated measurement cycles (100 – 200 s). Fig. 4(c) further presents the angular dependence of the THz emission obtained by rotating the WGP in key states 1 and 4 (low-amplitude, parallel alignment) and state 3 (high-amplitude, antiparallel alignment). The resulting polar plots reveal that the dominant THz polarization component resides in the *X*-plane of the laboratory frame.

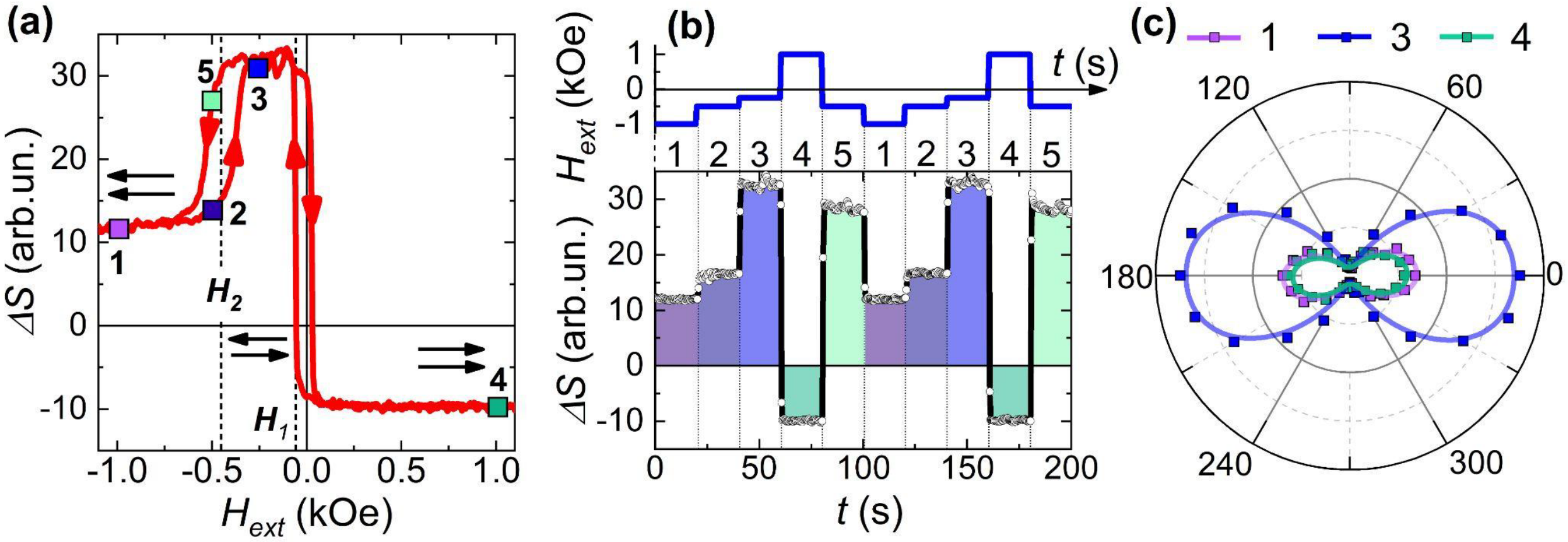

FIG 4. Magnetically induced modulation of THz-pulse amplitude in the spin-valve Co (1.8 nm)/Pt (3 nm)/Co (1.8 nm)/IrMn (5 nm). (a) TSM hysteresis loop measured at an optical pump fluence of 0.47 mJ/cm$^2$. Dots 1–5 indicate the values of the external magnetic field $H_{ext}$ used for subsequent analyses of THz amplitude modulation and polarization; the numbering reflects the sequential order in which $H_{ext}$ was swept. (b) Time-domain switching of the THz emission at points 1–5 recorded over two measurement cycles, with the temporal profile of $H_{ext}$ shown above. (c) Polar diagrams of the THz signal acquired while rotating the WGP for the low-amplitude states with parallel FM alignment (points 1 and 4) and the high-amplitude state (point 3): dots – experiment; solid red curves – fit (see Fig. S5, Supplemental materials [31]).

### B. Mechanism of THz amplitude modulation

Within the ISHE-based picture of THz emission, the observed amplitude modulation is interpreted as interference of waves generated by spin-to-charge conversion in the NM and antiferromagnetic layers, with constructive or destructive interference determined by the polarization of ultrafast spin currents in the FM layers. In turn, the spin-current polarization direction is determined by the initial orientation of the magnetization in each magnetic layer. To interpret the data quantitatively, we use a two-macrospin model (free and pinned Co) and compute equilibrium magnetic states as a function of the field by solving the Landau-Lifshitz-Gilbert (LLG) equation in the long-time limit. This quasi-static approach assumes that the external field variation is sufficiently slow compared to the characteristic magnetic relaxation times, allowing the system to remain in local thermodynamic equilibrium at each field value. The structure under investigation comprises two Co layers separated by the Pt spacer and capped by the antiferromagnetic IrMn layer. The Co layer adjacent to IrMn is magnetically hard (pinned), and its magnetization is denoted $M_2$. This pinned layer supports two spin-current channels: $j_{s2}$ (into Pt), which flows toward the Pt interface, and $j'_{s2}$ (into IrMn), which is converted within IrMn (Fig. 1). The lower Co layer is magnetically soft (free layer); its magnetization is designated $M_1$, and it generates the spin current $j_{s1}$(into Pt).

The free-energy density $F$ incorporates four additive contributions: (*i*) the Zeeman interaction between the net magnetization and the applied magnetic field $H_{ext}$; (*ii*) the dipolar (demagnetizing) energy, which minimizes the magnetostatic term by favoring in-plane alignment; (*iii*) the uniaxial magnetocrystalline anisotropy of each layer; and (*iv*) the bilinear exchange coupling between adjacent FM layers:

$$F_n = A(\vec{M}_n \cdot \vec{M}_{n+1}) - \frac{1}{2}\frac{H_{a,n}}{M_n}\left(\vec{M}_n \cdot \hat{n}\right)^2 + 2\pi\,(\vec{M}_n \cdot z)^2 - (\vec{M}_n \cdot \vec{H}_{ext}), \tag{1}$$

where $\vec{M}_n$ is the saturation magnetization of the $n$-th Co layer, $H_{an}$ is the uniaxial anisotropy field of the layer (Oe), $\hat{n}$ is the unit vector along the easy axis, $A$ is the normalized interlayer exchange coupling constant (dimensionless).

The magnetization dynamics is governed by a set of four coupled LLG differential equations expressed in spherical coordinates ($\theta_n$, $\varphi_n$):

$$\begin{gathered} \frac{\partial \theta_n}{\partial t} = \frac{\gamma M_n}{\Delta}\left[-\sin\theta_n \frac{\partial F_n}{\partial \varphi_n} - G\sin^2\theta_n \frac{\partial F_n}{\partial \theta_n}\right], \\ \frac{\partial \varphi_n}{\partial t} = \frac{\gamma M_n}{\Delta}\left[\sin\theta_n \frac{\partial F_n}{\partial \theta_n} - G\frac{\partial F_n}{\partial \varphi_n}\right], \end{gathered} \tag{2}$$

where $n = 1, 2$ (layer 1: free; layer 2: pinned), $\gamma$ the gyromagnetic ratio, $G$ the Gilbert damping constant, and $\Delta = M_n^2\left(1 + G^2\right)\sin^2\theta_n$. The Cartesian components of the magnetization vectors are:

$$M_n(t) = M_n\begin{pmatrix} \sin\theta_n \cos\varphi_n \\ \sin\theta_n \sin\varphi_n \\ \cos\theta_n \end{pmatrix} \tag{3}$$

Eqs. (2) and (3) are integrated numerically (fourth-order Runge-Kutta method) to obtain $M_n(t)$ for any applied field $H_{ext}$, with the integration time chosen to be long enough for the system to reach a quasi-equilibrium steady state. In this regime, the system relaxes to a state where the magnetization is determined by the balance between the applied field, magnetic anisotropy, and interlayer exchange coupling, with damping effects ensuring convergence to the equilibrium configuration. Numerical integration of the LLG system yields the time-dependent magnetization of each layer for any applied field (the inset of Fig. 5(a) shows the calculated total magnetization $M(H_{ext})$ for the free and pinned FM layers).

After obtaining $M_n(t)$ we model the emitted THz waveform under the assumption that the ISHE is the dominant conversion mechanism. For each layer the ISHE-induced electric field is evaluated as

$$\vec{E}_{THz} \propto \vec{j}_c \; = \; \Theta_{SH} \cdot \left(\vec{j}_s \times \vec{M}\right)$$

where $\Theta_{SH}$ is the spin-Hall angle of the conversion layer (Pt or IrMn), $\vec{M}$ is the magnetization of the Co layer, and $\vec{j}_s$ is the spin-current vector injected into the adjacent heavy-metal layer.

Spin currents generated in each Co layer are converted into charge currents predominantly in the Pt spacer and, to a lesser extent, in the IrMn cap. For the pinned Co/IrMn layer (top), the macrospin remains strictly in the film plane and is therefore written as $\vec{M}_2 = \left(0, M_{2y}, 0\right)$, where $M_{2y}$ is obtained from the LLG calculation. Two spin-current channels are assigned to this layer: $\vec{j}_{s2} = (0,0,-1)$ – the principal current injected into Pt, и $\vec{j}_{s2'} = (0,0,1)$ – the counter-flow converted inside IrMn. For the free Co layer (bottom): $\vec{M}_1 = \left(0, M_{1y}, 0\right)$, $\vec{j}_{s1} = (0,0,1)$, with $M_{1y}$ likewise provided by the LLG solver. At every bias field $H_{ext}$ we therefore evaluate three vector products that enter the inverse spin Hall expression and obtain the total THz field as the coherent sum:

$$E_{THz}(H_{ext}) \; \propto j_c = \sum_{i \in \{1,2,2'\}} \Theta_{SH}^{(i)}\left[\vec{j}_{si} \times \vec{M}_i\right],$$

Note that the two pinned-layer contributions (Pt and IrMn) share the same magnetization: $\vec{M}_2 = \vec{M}_{2'}$.

Each term isolates the contribution of a specific layer-interface pair; the resulting $E_{THz}(H_{ext})$ curve therefore disentangles the roles of the pinned and free Co layers as well as the separate spin-to-charge conversion processes in Pt and IrMn, furnishing a comprehensive picture of THz generation in the spin-valve. Fig. 5(a) contrasts calculated and measured $E_{THz}(H_{ext})$ loops for optical pump powers of 10 mW and 100 mW. The higher power merely improves the signal-to-noise ratio; the loop shape – and hence the underlying physics – remains unchanged, confirming the robustness of the model. All parameters obtained from the model are summarized in Table I. The IrMn spin Hall angle of 0.022 [33], which is 60% of the Pt value (0.04 [34]), validates the three-channel ISHE model when applied to polycrystalline IrMn/Pt heterostructures. Moreover, simple arithmetic shows that the resulting charge current amplitude from the pinned layer is $\vec{j}_{c2} + \vec{j}_{c2'} = 0.4\vec{j}_{c1}$.

The LLG calculations yield complete magnetization rotation angles for each FM layer. We used these results to analyze THz radiation polarization angles. Using the solution of Eq. (2) in spherical coordinates ($\theta_n$, $\varphi_n$), we computed the azimuthal angles $\varphi_1$ and $\varphi_2$ for the free and pinned layers ($\theta_n$ = 0 due to in-plane anisotropy). For each external field $H_{ext}$, we generated data arrays $\{H_{ext}, \varphi_1(H_{ext})\}$ and $\{H_{ext}, \varphi_2(H_{ext})\}$ (Fig. S8 in the Supplemental Material [31]). The observed divergence between $\varphi_1$ and $\varphi_2$ at identical $H_{ext}$ values explains the experimental THz polarization angle dependencies shown in Fig. 5(b).

We assume the experiment detects the vector sum of charge currents: $\vec{j}_{c1}(\varphi_1) + 0.4\vec{j}_{c1}(\varphi_2)$. The rotation angle of the THz polarization plane is then:

$$\Phi(H_{ext}) = \text{atan}\left(\frac{sin(\varphi_1) + 0.4sin(\varphi_2)}{cos(\varphi_1) + 0.4cos(\varphi_2)}\right)$$

where $\varphi_1 = \varphi_1(H_{ext})$ and $\varphi_2 = \varphi_2(H_{ext})$ are arrays from the LLG model.

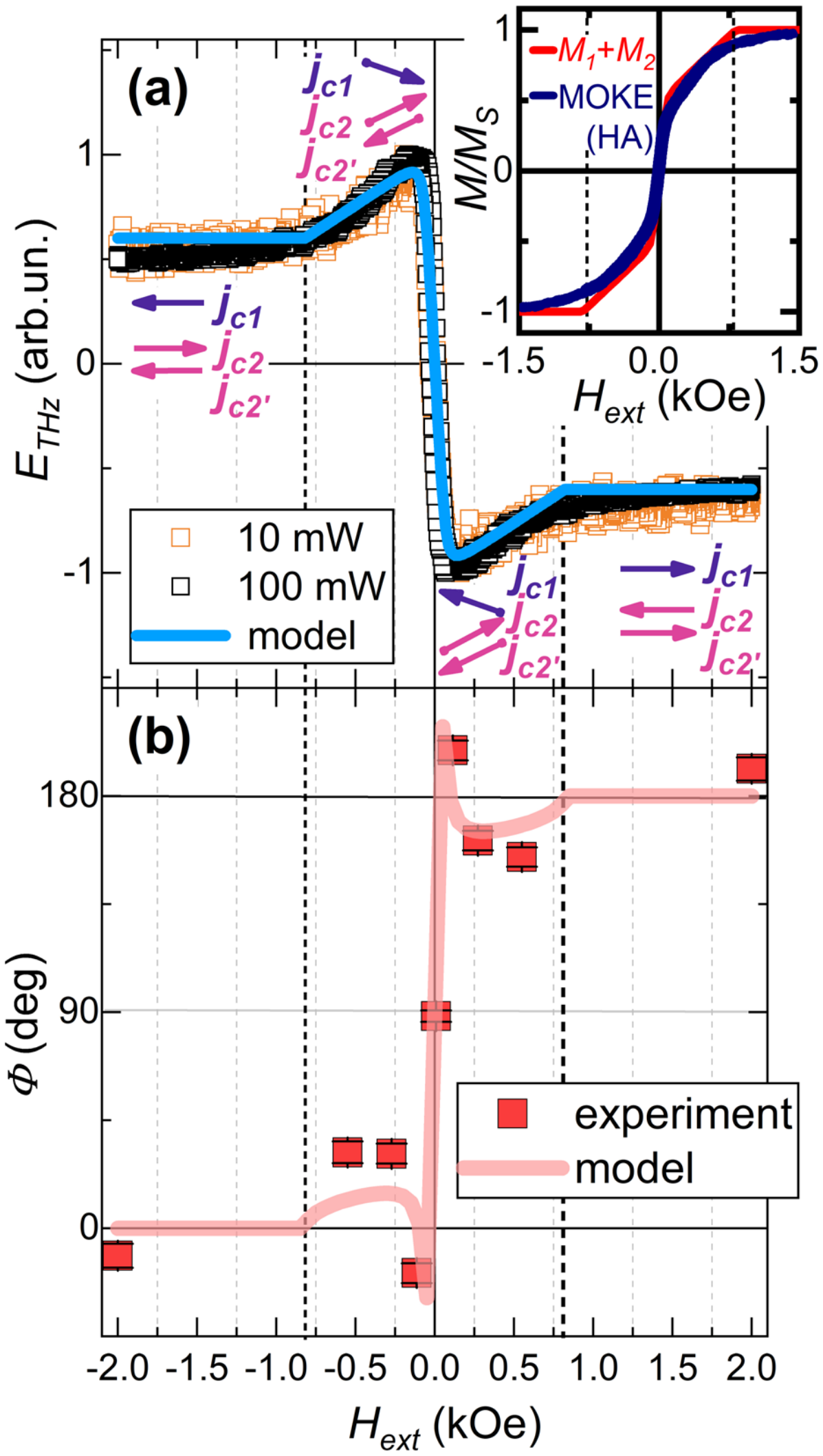

FIG 5. (a) Calculated (solid lines) and measured (symbols) $E_{THz}(H_{ext})$ loops for optical pump powers of 10 mW (blue) and 100 mW (red). The close agreement confirms that the two-macrospin ISHE model captures both the field dependence and the power-independent shape of the THz response. (b) Polarization angle $\Phi$ versus $H_{ext}$: orange squares – values extracted from fitting the measured angular diagrams $\Delta S(\alpha)$ in Fig. S7 (Supplemental Material [31]), where $\alpha$ is the angle between the WGP axis and the crystallographic direction [-110] of ZnTe; solid orange curve – the model.

Fig. 5(b) compares experimental polarization angles $\Phi$ (orange squares), extracted from angular diagram fits (Fig. S7 in the Supplemental Material [31]) with numerical predictions (solid curve). The good agreement confirms that THz polarization arises from the vector superposition of currents from both FM layers.

Table I: Model-derived parameters used to fit the experimental data.

| **Parameter** | Relative spin Hall angle, IrMn, ($\Theta_{SH}^{IrMn}$) | Spin Hall angle, Pt ($\Theta_{SH}^{Pt}$) | Anisotropy field, free layer ($H_{a1}$) | Anisotropy field, pinned layer ($H_{a2}$) | Gilbert damping ($G$) | Interlayer exchange constant ($A$) |
|---|---|---|---|---|---|---|
| **Value** | 0.6 $\Theta_{SH}^{Pt}$ | 1.0 (reference, applied to $j_{s1}$ and $j_{s2}$) | 0.05 kOe | 0.8 kOe | 0.007 [35] | -0.02 |

Fig. 6 (a) shows temporal waveforms of the terahertz signal for orthogonal components parallel to the laboratory $X$ and $Y$ axes (Fig. 1). We observe a nonzero time delay $\tau_d \approx 0.08$ ps between the main $S_Y$ and $S_X$ peaks, most pronounced at $H_{ext}$ = -0.55 kOe. This indicates the presence of a phase difference between the signals, which is estimated to be ~150° at 1 THz. We attribute this phase offset to the group delay during THz wave propagation in Pt and to an additional phase shift from multiple interferometric reflections in a Gires–Tournois-like multilayer resonator [24,36].

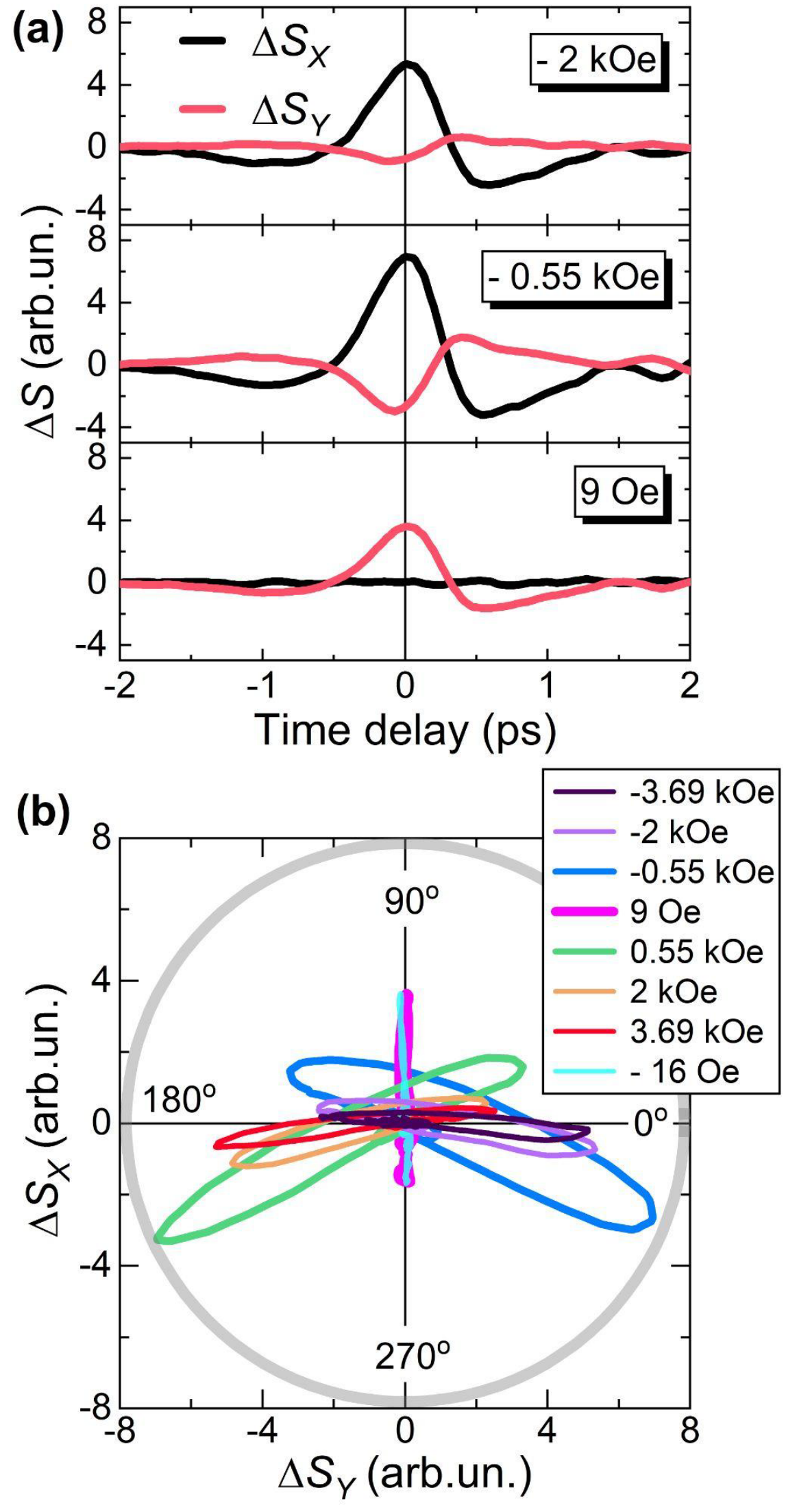

FIG 6. Ellipticity of THz radiation generated in the spin valve upon magnetization along HA. (a) temporal waveforms of orthogonal THz components $S_X$ and $S_Y$ at three specific values of $H_{ext}$: $i$) $0 < H_{ext} < H_{a1}$, $ii$) $H_{a1} < H_{ext} < H_{a2}$, and $iii$) $H_{ext} > H_{a2}$. (b) Lissajous curves of THz waves under different applied magnetic fields $H_{ext}$.

The Lissajous curves in Fig. 6(b) reveal elliptically polarized THz waves. Maximum ellipticity $\varepsilon$ is reached at $H_{ext}$ = -0.55 kOe, consistent with angular diagram fits (Fig. S7 in the Supplemental Material [31]). Linear THz polarization along HA appears only near zero field (9 Oe and − 16 Oe). Under these conditions, angles $\varphi_1$ and $\varphi_2$, – which determine charge current orientation in the NM spacer – coincide (Fig. S8 in the Supplemental Material [31]).

Notably, at high fields $H_{ext}$ = ± 3.7 kOe (significantly exceeding the pinned layer anisotropy field $H_{a2}$), a small ellipticity (~ 0.15) persists (Fig. 6(b)), indicating residual domain structure even where MOKE suggests macroscopic saturation (Fig. 2(b)). THz spectroscopy demonstrates enhanced sensitivity to the spin valve magnetic structure. External fields from 2 kOe to 3.7 kOe prove insufficient for complete domain moment alignment (Fig. S9 in the Supplemental Material [31]).

## IV. CONCLUSIONS

In this work, we realized an ultrathin Co/Pt/Co/IrMn spin-valve with a total metal thickness of 11.6 nm that emits THz radiation under femtosecond optical pumping and supports efficient amplitude modulation with a depth of 65%. Compared with a Co (2 nm)/Pt (2 nm) reference THz emitter, the THz field amplitude in the high-amplitude state is lower by approximately a factor of five. Experiment and modeling consistently indicate three spin-to-charge conversion channels: two in Pt (contributions from the free and pinned Co layers) and an additional one in IrMn. This third channel accounts for the incomplete suppression of the signal in the parallel configuration. A macrospin LLG model, combined with an ISHE-based description, reproduces the dependence $E_{THz}(H_{ext})$, separates the layer-specific contributions, and indicates $\Theta_{SH}^{IrMn} \approx 0.6\ \Theta_{SH}^{Pt}$. Field-steered polarization, including emergent ellipticity in the hard-axis geometry, arises naturally from the vector superposition of layer-resolved charge currents with a small intercomponent phase delay. Collectively, these results define a compact spin-valve architecture that combines low-field, multilevel THz amplitude control with polarization tunability, offering a practical route toward on-chip THz modulators and nonvolatile memory-like functional elements, and establishing THz spintronic magnetometry as a sensitive probe of exchange-biased heterostructures.

**ACKNOWLEDGMENT**

The work was supported by the Russian Science Foundation under Grant No. 23-19-00849 (https://rscf.ru/en/project/23-19-00849/). The development and numerical solution of the two-macrospin LLG model, including all micromagnetic simulations, were carried out with the support of the were funded by The Ministry of Science and Higher Education (Contract No. FSFZ-2025-0002).

AUTHOR CONTRIBUTION

Arseniy M. Buryakov: conceptualization (lead); modeling(lead); formal analysis (lead); validation (equal); writing – original draft (lead); writing – review & editing (equal). Anastasia V. Gorbatova: investigation (lead); validation (lead); writing – original draft (lead); visualization (lead); Pavel Y. Avdeev: investigation (lead); data curation (supporting); Igor Yu. Pashen'kin: investigation (lead); Maksim V. Sapozhnikov: writing – review & editing (lead). Alexey A. Klimov: Investigation(equal); Elena D. Mishina: writing – review & editing (lead). Vladimir L. Preobrazhensky: supervision (equal); modeling (equal); writing – review & editing (lead).

CONFLICT OF INTEREST

The authors declare no competing financial interest.

DATA AVAILABILITY

The data that support the findings of this article are openly available [31].